\documentclass[aps,prd]{revtex4-2}
\usepackage[british]{babel}
\usepackage{csquotes}

\usepackage{amsmath}
\usepackage{graphicx}
\usepackage[colorlinks=true, allcolors=blue]{hyperref}
\usepackage{xcolor}
\usepackage{slashed}
\providecommand{\Tr}{\operatorname{Tr}}
\providecommand{\tcr}[1]{#1}
\providecommand{\tcb}[1]{}
\graphicspath{{./}{Figures/}}
\begin{document}
\title{Gribov copies in the quark propagator}
\author{Gerhard Kalusche}
\author{Dale Lawlor}
\author{Jon-Ivar Skullerud}
\affiliation{Department of Physics, National University of Ireland Maynooth, Maynooth, Co Kildare, Ireland}


\begin{abstract}
We study the impact of Gribov copies on the quark propagator in
lattice 2-colour QCD.  We find that the Gribov noise is comparable to
the gauge noise for smaller volumes but becomes less significant for
larger spatial volumes.  The Gribov noise in the quark propagator is
found to be comparable to, but smaller than in the gluon propagator on
the same ensembles.  No correlation is found between the values of the
wave function $Z(p)$ and the value of the gauge fixing functional, nor
between the two form factors.  A very mild negative correlation was found
between the value of the quark mass function $M(p)$ and the gauge
fixing functional.
\end{abstract}
\maketitle
\section{Introduction}

Quantum Chromodynamics (QCD) describes the strong interaction as arising from the fundamental quark and gluon fields, which transform non-trivially under the SU(3) gauge group.  Although all actually observable quantities arising out of the theory are gauge invariant, the fundamental Green's functions (propagators and vertices) are not, and gauge fixing is required to compute them.  In a nonabelian gauge theory this gauge fixing is not unique, leading to the appearance of Gribov copies \cite{Gribov:1977wm,Singer:1978dk,DellAntonio:1991mms,Sobreiro:2005ec}, that is, multiple gauge field configurations that satisfy the same gauge-fixing condition, but still belong to the same gauge orbit.

The problem is ameliorated by restricting the configurations to the first Gribov region, which is the set of minima of the gauge fixing functional $||A||^2$, and is, equivalently, characterised by a positive definite Faddeev--Popov operator.  This condition is implemented, variously, by the introduction of a horizon condition in the Gribov--Zwanziger (GZ) and refined Gribov--Zwanziger (RGZ) actions \cite{Vandersickel:2012tz,Dudal:2008sp}; as boundary condition in functional equations including Dyson--Schwinger equations (DSEs) and the functional renormalisation group (FRG) \cite{Fischer:2006ub,Ferreira:2023fva,Gies:2006wv,Dupuis:2020fhh}; or by numerically minimising the gauge fixing functional in lattice gauge theory.  This, however, does not eliminate the problem since the first Gribov region still contains multiple copies and the different approaches sample these copies in an uncontrolled manner, making any comparison between their results problematic.

Attempts to select a unique representative of each gauge orbit are hampered by several issues.  There are gauges such as axial gauges and the lattice Laplacian gauge which are free of Gribov copies, but the former violate Lorentz covariance, while the latter has no known continuum counterpart.  These are manifestations of a theorem by Singer \cite{Singer:1978dk}, that no unique and continuous gauge fixing condition exists.  Finding the absolute minimum of the gauge fixing functional (known as the fundamental modular region, FMR) numerically is unfeasible since this is an NP-hard problem and the number of Gribov copies grows exponentially with the volume.  The same goes for any other numerical prescription.
It has been conjectured \cite{Greensite:2004ur} that in the infinite-volume limit the important configurations lie on the common boundary of the first Gribov region and the FMR, and if this holds then randomly selecting Gribov copies within the first Gribov region would be sufficient.

In the absence of any solution to the Gribov problem, it is important to establish whether and to what extent the fundamental propagators and vertices of the theory, and by implication any observables that may be derived from these, are sensitive to the ambiguity.  There have been a number of lattice studies investigating the impact of Gribov copies on gluon and ghost propagators in pure Yang--Mills theories
\cite{Silva:2007tt,Boucaud:2011ug,Bornyakov:2011fn,Sternbeck:2012mf,Maas:2015nva,Maas:2017csm} as well as in gauge--Higgs theories \cite{Maas:2010nc} and theories with adjoint fermions \cite{Maas:2011jf}.  In brief, the impact on the gluon propagator is found to be modest (up to 10\% in 4 dimensions), and only at the lowest momenta, decreasing with increasing volume.  For the ghost propagator, a more significant effect is found, with differences up to a factor 3 over a wider range of momenta.  It is worth noting that these studies have been carried out using both the SU(2) and SU(3) gauge group, and no qualitative difference between the two gauge groups has been found.

In contrast, to our knowledge, no studies of the effects of Gribov copies on the third fundamental two-point function of QCD, the quark propagator, have as yet been carried out, and hence all lattice and functional studies of the quark propagator are in principle afflicted by unquantified systematic uncertainties arising from the Gribov ambiguity.  This investigation is an attempt to rectify this situation.  It has been performed in the computationally simpler SU(2) gauge group, which shares the salient features of quark confinement and dynamical chiral symmetry breaking with real QCD.

The remainder of this paper is structured as follows.  In section~\ref{sec:technical} we present our computational setup, including the gauge and fermion action and parameters, the gauge fixing procedure, and the definition of the quark propagator form factors we will be studying.  In section~\ref{sec:results} we present our results for the quark propagator form factors and their variation with gauge copies and gauge configurations.  Finally, in sec.~\ref{sec:conclusion} we summarise our findings and discuss some open questions.

\section{Computational setup}
\label{sec:technical}

We have simulated QCD with gauge group SU(2) (QC$_2$D) using a Wilson gauge action and $N_f=2$ Wilson fermions.  This is part of an ensemble used for studying QC$_2$D at high density \cite{Hands:2006ve,Hands:2010gd,Cotter:2012mb,Boz:2013rca,Boz:2019enj,Lawlor:2022hfi}.  The parameters used in this study correspond to the `coarse' ensemble in \cite{Boz:2019enj}. Three different lattice volumes have been used to assess temperature and volume effects.  The parameters are given in table~\ref{tab:params}.

\begin{table}
\centering
\begin{tabular}{ll|ccc|rrcc}
$\beta$ & $\kappa$ & $a$ (fm) & $m_\pi/m_\rho$ & $m_q$ (MeV) & $N_s$ &$N_\tau$
& $V\, (\mathrm{fm}^3)$ & $T$ (MeV) \\\hline
1.9 & 0.1680 & 0.178(6) & 0.805(9) & 56 & 24 & 12 & 78 & 94 \\
1.9 & 0.1680 & 0.178(6) & 0.805(9) & 56 & 16 & 12 & 23 & 94 \\
1.9 & 0.1680 & 0.178(6) & 0.805(9) & 56 & 16 & 24 & 23 & 47 
\end{tabular}
\caption{\label{tab:params}Simulation parameters: gauge coupling $\beta$, hopping parameter $\kappa$, lattice spacing $a$, pseudoscalar-to-vector mass ratio $m_\pi/m_\rho$, subtracted bare quark mass $m_q$, spatial and temporal extent $N_s, N_\tau$, lattice volume $V$ and temperature $T$.}
\end{table}

The lattice configurations have been fixed to Landau gauge by maximising the functional 
\begin{equation}
  F[U;g] = \sum_{x,\mu}U^g_\mu(x)
  = g(x)U^g_\mu(x)g^\dagger(x+\hat{\mu}) \label{eq:functional}
\end{equation}
using a standard overrelaxation algorithm, with the convergence criterion \begin{equation}
   \sum_{x,\mu,a}[A^a_\mu(x+\mu)-A^a_\mu(x)]^2<10^{-12}\,,
\end{equation}
where the gauge potential $A_\mu^a(x)$ is defined by
\begin{equation}
A_\mu^a(x) = \frac{1}{2i}\operatorname{tr}\Big(\tau^a[U_\mu(x)-U_\mu^\dagger(x)]\Big)\,,
\end{equation}
and $\tau^a$ are the Pauli matrices.  Different Gribov copies have been obtained by repeating this procedure $N_{\text{copy}}=100$ times after a random gauge transformation.  This has been done for 5 different gauge configurations (separated by 20 Hybrid Monte Carlo trajectories) for each of the three lattice volumes.

The quark propagator $S(x,y;U)$ on a single configuration $U$ is computed by inverting the Wilson fermion matrix,
\begin{gather}
    S(x,y;U) = (M_{W}[U])^{-1}_{xy} \label{eq:Wilson-inverse}\\
    M_{W}[U]_{xy} = (m_0+4)\delta_{xy}
    - \frac{1}{2}\sum_{\mu=\pm1}^{\pm4}(1-\gamma_\mu)U_{\mu}(x)\delta_{x+\hat{\mu},y} \label{eq:Wilsonmatrix}
\end{gather}
Following the gauge fixing, the gauge transformed propagator is given by
\begin{equation}
    S(x,y;U^g) = g(x)S(x,y;U)g^\dagger(y)\,.
\end{equation}
The inversion has been carried out using a single point source at the origin, and the gauge fixed propagator has been Fourier transformed to momentum space.  The discrete lattice momenta entering into the Fourier transform are given by
\begin{align}
    p_i &= \frac{2\pi n_i}{L_i}\quad (i=1,2,3)\, 
    & p_4 &= \frac{(2n_4+1)\pi}{L_\tau}\,. \label{eq:fourier-momenta}
\end{align}
In the continuum, the Euclidean-space quark propagator can be written in terms of three form factors $A, B$, and $C$,
\begin{equation}
    S^{-1}(p) = i\vec{\gamma}\cdot\vec{p}A(\vec{p}^2,p_t)
    + i\gamma_4\omega C(\vec{p}^2,p_4) + B(\vec{p}^2,p_4)\,,
\end{equation}
where $\omega=p_4-i\mu$ and $\mu$ is the quark chemical potential.  At zero temperature and chemical potential, $A$ and $C$ become degenerate and the propagator can be written as
\begin{equation}
    S(p) = \frac{1}{i\slashed{p}A(p^2)+B(p^)}
    = \frac{Z(p^2)}{i\slashed{p}+M(p^2)}\,. \label{eq:quark-prop-cont-T0}
\end{equation}
The ensembles used in this study are all at $\mu=0$, with one at effectively zero temperature and the other two at $T=94\,$MeV, which is still well below the deconfinement phase transition \cite{Boz:2013rca}.  As our focus is on the impact of Gribov copies, we will here ignore any medium effects and take \eqref{eq:quark-prop-cont-T0} to describe the continuum quark propagator for all ensembles.

The tree-level lattice fermion propagator with the Wilson action \eqref{eq:Wilsonmatrix} is given by
\begin{equation}
    S(p) = \frac{1}{i\slashed{K}(p) + m_0 + \frac{a}{2}Q^2(p)}\,,
    \label{eq:Wilson-treelevel}
\end{equation}
where we have introduced the lattice momentum variables
\begin{align}
    K_\mu(p) &= \frac{1}{a}\sin(ap_\mu)\,,
    & Q_\mu(p) &= \frac{2}{a}\sin\Big(\frac{ap_\mu}{2}\Big)\,.
\end{align}
Using this as a guide, we define the ``tree-level corrected'' lattice form factors $Z(p)$ and $M(p)$ as
\begin{align}
    Z(p) &= \frac{\mathcal{A}(p)}{K^2(p)\mathcal{A}^2(p)+\mathcal{B}^(p)}\,,
    \label{eq:Z-lat-corr}\\
    M(p) &= \frac{\mathcal{B}(p)}{\mathcal{A}(p)}\frac{m}{m+\frac{a}{2}Q^2(p)}\,,
    \label{eq:M-lat-corr}
\intertext{where}
\mathcal{A}(p) &= \frac{i}{4K^2(p)}\Tr\big[\slashed{K}(p)S(p)\big]\,,\\
\mathcal{B}(p) &= \frac{1}{4}\Tr\big(S(p)\big)\,,
\end{align}
and $m$ is the subtracted bare quark mass, $m=m_0-m_c$ with the critical quark mass $m_c$ defined as where the pion mass vanishes.
With these definitions we have $Z(p)=1$ and $M(p)=m_0$ for noninteracting fermions, and due to asymptotic freedom we expect that $Z(p)\to1$ and $M(p)\to m_{\text{cur}}$ in the ultraviolet, where $m_{\text{cur}}$ is the renormalisation group invariant current quark mass. It should be noted that the tree-level correction (the second term) in \eqref{eq:M-lat-corr} only gives a constant contribution which is identical for all gauge configurations and Gribov copies, and therefore has no impact on how $M(p)$ varies between different Gribov copies, which is the main concern of this study.

\section{Results}
\label{sec:results}

\begin{figure}
\centering
\includegraphics[width=0.48\linewidth]{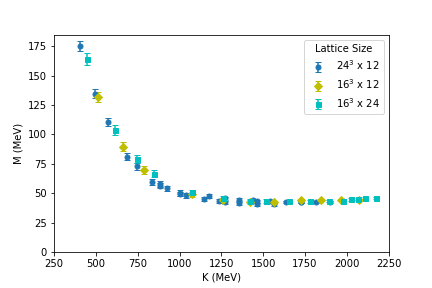}
\includegraphics[width=0.48\linewidth]{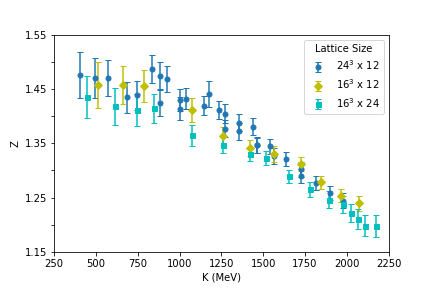}
\caption{The tree-level corrected mass function $M(p)$ (left) and wave function $Z(p)$ (right) versus four-momentum $K(p)$ for all three lattice volumes.  The data have been cylinder cut to reduce lattice artefacts, see text for details.}
\label{fig:M-Z-vs-p}
\end{figure}

In figure~\ref{fig:M-Z-vs-p} we show the mass function $M(p)$ and wave
function $Z(p)$ as functions of momentum $K(p)$ for each of our three
ensembles.  The data have been obtained from 50 configurations for
each ensemble.  To reduce lattice artefacts, a cylinder cut \cite{Leinweber:1998im} has been imposed, selecting momenta within a radius $a\Delta p<\frac{8}{\pi}\approx2.5$ from the 4-diagonal.  

The mass function exhibits a clear infrared enhancement signalling dynamical chiral symmetry breaking.  In the ultraviolet it approaches a constant value of about 50 MeV, which is comparable to the subtracted bare quark mass for these ensembles, $m = 56$ MeV.  The wave function $Z(p)$ is enhanced in the infrared, unlike what has been found in SU(3) with various fermion formulations \cite{Oliveira:2018lln,Bowman:2005vx,Kamleh:2004aw,Schrock:2011hq,Blossier:2010vt,Virgili:2022wfx}.  We do not know if this difference is a feature of 2-colour QCD or an effect of using unimproved Wilson fermions on coarse lattices; further studies with finer lattices would be required to determine this.

We do not find any strong temperature or finite-volume effects in either quantity.  It is worth noting that the hotter $N_\tau=12$ lattices still correspond to a temperature well below the deconfinement crossover \cite{Boz:2013cra}, so the absence of any significant temperature dependence is not surprising.

\begin{figure}
\centering
\includegraphics[width=0.45\linewidth]{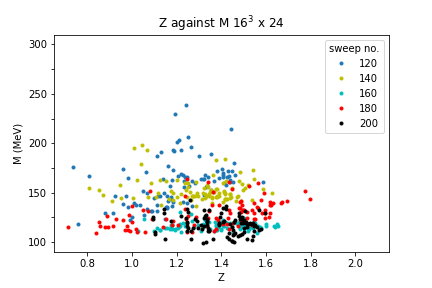}
\includegraphics[width=0.45\linewidth]{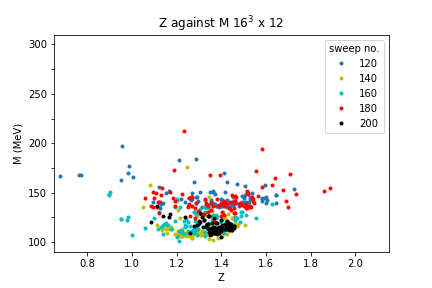}
\includegraphics[width=0.45\linewidth]{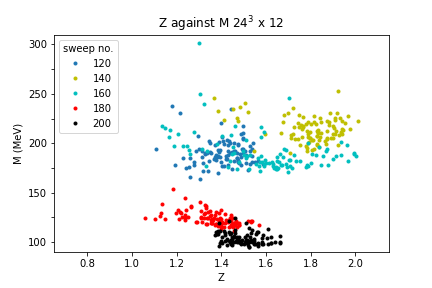}
\caption{\label{fig:MvsZ}The quark mass function $M(p)$ and wave function $Z(p)$ at four-momentum $K(p)\approx500\,$MeV, for the three lattice volumes.  Different colours represent different gauge configurations; each point represents a distinct gauge copy.}
\end{figure}

We now turn to how the quark propagator varies between different Gribov copies.
In Fig.~\ref{fig:MvsZ} we show $M(p)$ and $Z(p)$ from each of the 100
Gribov copies on each of the 5 configurations studied.  \tcr{The means
  and standard deviations are tabulated, along with those of the gluon
  propagator and the gauge fixing functional, in
  Table~\ref{tab:dispersion}.}  In order to compare results between
the different lattice volumes, we use (approximately) the same
physical momentum value ($K(p)\approx500\,$MeV) for all three
lattices \footnote{Specifically, the four-momentum values are $K(p)=448$\,MeV, 511\,MeV and 496\,MeV for the $16^3\times24, 16^3\times12$ and $24^3\times12$ lattices respectively.  For the gluon propagator we will use four-momenta 432\,MeV, 432\,MeV and 501\,MeV for the same three lattices.}.  We see that for the smaller spatial volume, the Gribov noise is comparable to the gauge noise, while for the larger $24^3$ volume it is significantly smaller (the sets of points with different colours do not overlap much).  For the larger volume, the Gribov noise appears to be more significant for $Z(p)$ than for $M(p)$.  We do not find any correlation between the two form factors.

\begin{table}
\centering
\begin{tabular}{c|c|c|c|c|c}
lattice & config. & $F(\sigma_F)$ & $aM(\sigma_M)$ & $Z(\sigma_Z)$ & $a^{-2}D(\sigma_D)$ \\\hline
$16^3\times24$ & 120 & 0.81543(15) & 0.144(21) & 1.20(17) & 5.23(77) \\
 & 140 & 0.81553(14) & 0.138(11) & 1.30(17) & 5.54(104) \\
 & 160 & 0.81461(16) & 0.105(4) & 1.38(15) & 5.04(91) \\
 & 180 & 0.81466(14) & 0.116(12) & 1.34(25) & 5.95(81) \\
 & 200 & 0.81421(10) & 0.106(9) & 1.39(12) & 5.77(103) \\\hline
$16^3\times12$ & 120 & 0.81353(16) & 0.132(11) & 1.36(19) & 5.94(84) \\
 & 140 & 0.81456(18) & 0.105(11) & 1.30(10) & 4.64(89) \\
 & 160 & 0.81538(17) & 0.109(9) & 1.27(14) & 6.42(91) \\
 & 180 & 0.81656(21) & 0.129(12) & 1.39(17) & 5.42(88) \\
 & 200 & 0.81371(16) & 0.105(5) & 1.37(7) & 5.81(97) \\\hline
$24^3\times12$ & 120 & 0.81429(14) & 0.172(12) & 1.41(9) & 4.97(71) \\
 & 140 & 0.81501(9) & 0.193(11) & 1.79(13) & 4.91(84) \\
 & 160 & 0.81527(11) & 0.172(17) & 1.60(21) & 5.35(81) \\
 & 180 & 0.81483(10) & 0.112(6) & 1.36(10) & 5.03(72) \\
 & 200 & 0.81516(9) & 0.094(5) & 1.49(7) & 5.00(70) 
\end{tabular}
\caption{\label{tab:dispersion}Means and standard deviations (in lattice units) for the gauge fixing functional $F$ and the form factors $M, Z$ and $D$ at momentum $K(p)\approx500\,$MeV, for each configuration and lattice volume.}
\end{table}

\begin{figure}
\centering
\includegraphics[width=0.45\linewidth]{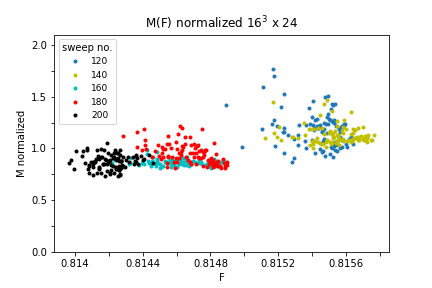}
\includegraphics[width=0.45\linewidth]{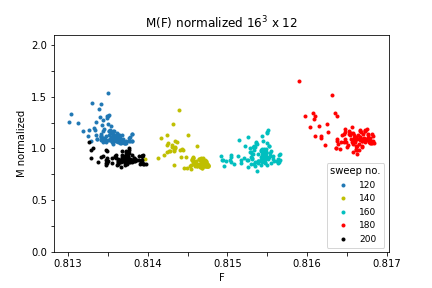}
\includegraphics[width=0.45\linewidth]{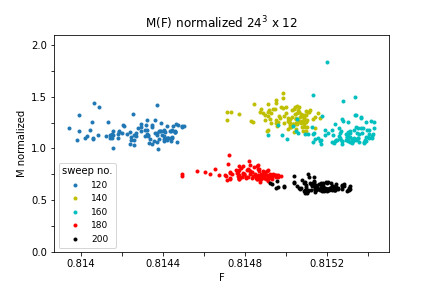}\\
\caption{\label{fig:Mnormalised}The quark mass function $M(p)$ at four-momentum $K(p)\approx500\,$MeV, for each of the three lattice volumes, versus the gauge fixing functional $F$. Different colours represent different gauge configurations, while each dot represents a Gribov copy.}
\end{figure}
\begin{figure}
\centering
\includegraphics[width=0.45\linewidth]{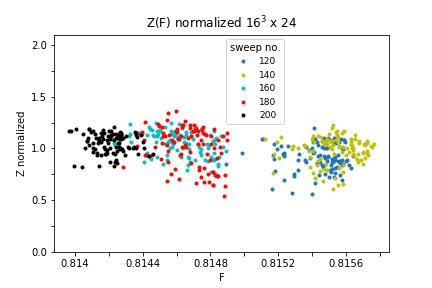}
\includegraphics[width=0.45\linewidth]{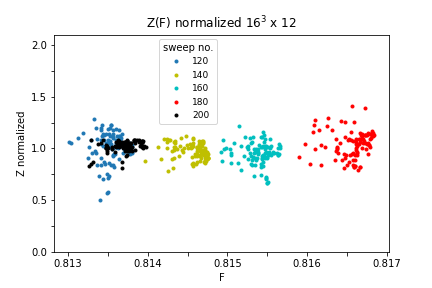}
\includegraphics[width=0.45\linewidth]{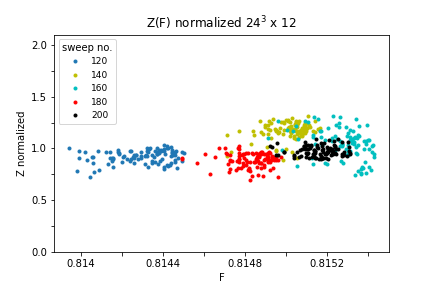}\\
\caption{\label{fig:Znormalised}The quark wave function $Z(p)$ at four-momentum $K(p)\approx500\,$MeV, for each of the three lattice volumes, versus $F$.}
\end{figure}
\begin{figure}
\centering
\includegraphics[width=0.45\linewidth]{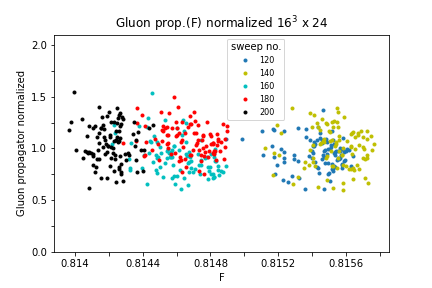}
\includegraphics[width=0.45\linewidth]{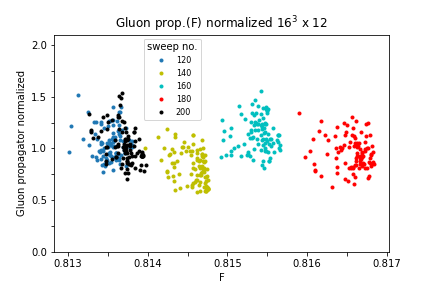}
\includegraphics[width=0.45\linewidth]{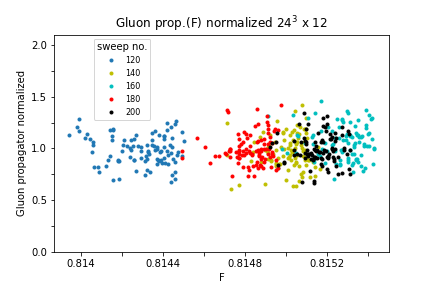}
\caption{\label{fig:Dnormalised}The gluon propagator at four-momentum $K(p)\approx500\,$MeV, for each of the three lattice volumes, versus $F$.}
\end{figure}

In order to assess the significance of the Gribov noise and to make a comparison between the two form factors, we now normalise the form factors by dividing the values for each gauge configuration and gauge copy by the ensemble average.  We also perform the same analysis of the gluon propagator, so that we can compare the importance of Gribov copies for the quark and gluon propagators.  The results are shown in Figs.  \ref{fig:Mnormalised}, \ref{fig:Znormalised} and \ref{fig:Dnormalised}, 
where the values of the form factors are shown as functions of the gauge fixing functional $F$ given in \eqref{eq:functional}.


We can see that the Gribov noise is somewhat smaller for the quark propagator than for the gluon propagator.  This is the case for all three lattice volumes.  We see, as before, that the Gribov noise for the quark propagator is reduced for the larger spatial volume, in clear contrast to what we can see for the gluon propagator.  Indeed, for the smaller lattices the Gribov noise is comparable to or even larger than the gauge noise, while it is smaller than the gauge noise for the larger lattice.  We see no clear evidence of any temperature dependence, as the results for the $16^3\times24$ and $16^3\times12$ lattices are very similar.

\tcb{OLD TEXT: We find no evidence of any correlation between the values of value of the quark or gluon propagator and the gauge fixing functional.  On the face of it, this would suggest that the fundamental modular region, defined as the absolute maximum of $F$ for each gauge orbit, would on average give the same result as picking a random copy.}

\begin{table}
\centering
\begin{tabular}{c|c|c|c}
variable & $16^3\times24$ & $16^3\times12$ & $24^3\times12$ \\\hline
$M(F)$ & -0.199(86) & -0.349(135) & -0.160(38) \\
$Z(F)$ & -0.143(85) & 0.039(98) & -0.040(116) \\
$D(F)$ & -0.171(62) & -0.208(89) & 0.000(40)  
\end{tabular}
\caption{\label{tab:correlation}Correlation coefficients between the form factors $M, Z$ and $D$ and the gauge fixing functional $F$ for each lattice.  The coefficients are calculated separately for each configuration and averaged; the uncertainties denote the spread between configurations.}
\end{table}

\tcr{From figures \ref{fig:Mnormalised}, \ref{fig:Znormalised} and \ref{fig:Dnormalised} we do not see any obvious correlation between the values of the quark or gluon propagator and the gauge fixing functional.  For a quantitative assessment, we have performed linear fits to each form factor as functions of the gauge fixing functional $F$.  The fits have been performed separately for each configuration in order to isolate the effect of the Gribov noise, with the resulting coefficients averaged over configurations.  The average correlation coefficients and their spread between configurations are shown in table~\ref{tab:correlation}. 
We find a mild negative correlation between the mass function $M$ and the gauge fixing functional, but no significant correlation for the wave function $Z$.  For the gluon propagator there is a mild negative correlation at smaller volumes, but no significant effect for the larger spatial volume.  Taken at face value, these results suggest that the fundamental modular region, defined as the absolute maximum of $F$ for each gauge orbit, would on average give the same result as picking a random copy, except possibly for the mass function.}

\section{Discussion and conclusions}
\label{sec:conclusion}

We have performed what is, as far as we know, the first systematic study of the effect of Gribov copies on the quark propagator in lattice QCD.  We find that, although the effect is visible and comparable to the gauge noise for small volumes, it is significantly reduced as the lattice volume is increased, in contrast to the situation for the gluon propagator.  This suggests that Gribov copies are not a significant confounding factor in lattice studies of the quark propagator or in comparing lattice to results from Dyson--Schwinger equations or the functional renormalisation group, where the sampling of the first Gribov region is not yet understood.

This study was carried out on relatively small lattices, so for a full understanding of the volume effects it would need to be repeated for larger lattice volumes.  Although our initial findings suggest that the effect of Gribov copies is reduced with increasing volume, this may be counteracted by the known exponential increase in the number of Gribov copies \cite{Hughes:2012hg,Mehta:2014jla}.  In particular, it is possible that outliers with extreme values for $Z$ and/or $M$ may become more prominent with increasing volume.  However, although the number of Gribov copies increases, the distribution of the gauge fixing functional $F[U]$ becomes narrower with increasing volume \cite{Mehta:2014jla}, something that can also be seen in Fig.~\ref{fig:Mnormalised},~\ref{fig:Znormalised},~\ref{fig:Dnormalised} in this paper.  Our observed narrowing of the distribution of $M$ and $Z$ may be taken as an indication of the same effect.

Another open question is finite lattice spacing effects, as this study was performed on a single, coarse lattice.  Past studies of the gluon and ghost propagators \cite{Maas:2017csm} have found that discretisation effects are rather small and suggest that the lattice spacing used in this study is at the edge of the region where they become insignificant.  However, this should be verified with a future study on a finer lattice with a matching volume.

This study has also been performed with a single, relatively large quark mass.  It therefore does not address the question of whether Gribov copy effects may become prominent for physical light quarks.  This is an important outstanding issue for future investigations.


\section*{Software and data}

The gauge configurations were generated with the {\tt su2hmc} code \cite{lawlor_dale_2022_7164407},
whereas the analysis code and the data presented in this paper are available at \cite{zenodo:gribov-quark}.

\section*{Author contributions}

\begin{description}
\item[GK] Secondary data production, analysis code, data analysis and plots.
\item[DL] Simulation code and primary data production.
\item[JIS] Primary data production, gauge fixing and quark propagator code, manuscript production.
\end{description}

\section*{Acknowledgements}

GK acknowledges support from a Maynooth University SPUR fellowship. 
DL acknowledges support from the National University of Ireland, Maynooth’s John and Pat Hume Scholarship.  JIS acknowledges support from Science Foundation Ireland grant 11-RFP.1-PHY3193.

The authors wish to acknowledge the Irish Centre for High-End Computing (ICHEC) for the provision of computational
facilities and support.
%
This work was in part performed using the DiRAC Data Intensive service at Leicester, operated by the University of Leicester IT
Services, which forms part of the STFC DiRAC HPC Facility (\url{www.dirac.ac.uk}). The equipment was funded by BEIS capital
funding via STFC capital grants ST/K000373/1 and ST/R002363/1 and STFC DiRAC Operations grant ST/R001014/1. DiRAC is
part of the National e-Infrastructure.

\bibliography{gluon,density,gribov-prd-resubmitNotes}

\end{document}